\documentclass[preprint,prb,aps,showpacs]{revtex4}%
\usepackage{amsfonts}
\usepackage{amsmath}
\usepackage{amssymb}
\usepackage{graphics}
\usepackage{graphicx}%
\begin{document}
\title{Theory of emission from an active photonic lattice}
\author{Weng W. Chow}
\affiliation{{Sandia National Laboratories, Albuquerque, NM 87185-0601}}
\date{July 12, 2005}


\begin{abstract}
The emission from a radiating source embedded in a photonic lattice is
calculated. The analysis considers the photonic lattice and free space as a
combined system. Furthermore, the radiating source and electromagnetic field
are quantized. Results show the deviation of the photonic lattice spectrum
from the blackbody distribution, with intracavity emission suppressed at
certain frequencies and enhanced at others. In the presence of rapid
population relaxation, where the photonic lattice and blackbody populations
are described by the same equilibrium distribution, it is found that the
enhancement does not result in output intensity exceeding that of the
blackbody at the same frequency. However, for slow population relaxation, the
photonic lattice population has a greater tendency to deviate from thermal
equilibrium, resulting in output intensities exceeding those of the blackbody,
even for identically pumped structures.

\end{abstract}
\maketitle

\section{\bigskip Introduction}

One of the many novel optical phenomena exhibited by photonic lattices is the
modification of spontaneous emission properties. \cite{pc1,pc2} A photonic
lattice can funnel radiation into narrow energy bands, where exceedingly high
intensities at photonic lattice bandedges have been predicted theoretically
and observed experimentally. \cite{piqeat,lin1,fleming,li} A question is
whether the peak intensities exceed those of a blackbody under similar
experimental conditions. \cite{dowling,lin,lin2,narayanaswamy,trupke,luo} The
answer is important for scientific understanding and can impact the
development of new light sources.

It is generally agreed that the higher photonic-lattice density of states will
increase intracavity intensity. The debate concerns the output intensity in
comparison with that of a blackbody. Arriving at an answer is difficult
experimentally because it is difficult to ensure that the comparison is made
under similar conditions. Theoretically, the difficulty lies with the
treatment of the matter and optical aspects of the problem. \cite{siegman} The
derivation of the matter equations requires knowledge of the normal modes of
the optical structure, preferably in the form of an orthonormal basis.
However, such a basis set is not rigorously defined for a finite photonic
lattice with outcoupling loss. This problem occurs also in laser theory, where
one usually begins with the Fox-Li modes for a Fabry-Perot cavity with
perfectly reflecting mirrors, and introduces a loss mechanism to represent the
outcoupling. \cite{lamb} Such a phenomenological approach is inadequate for
the present problem because of the inconsistency arising from separating the
treatments of the eigenmode problem and the outcoupling effects.

The approach taken in this paper\ considers the photonic lattice and free
space outside of the photonic lattice as one combined system (see Fig. 1). We
follow the method of an earlier paper on the linewidth of a Fabry-Periot laser
\cite{lang} in representing free space by a very large cavity. The photonic
lattice is approximated by a series of semitransparent interfaces. We begin
with discussing the one-dimensional geometry, \cite{dowling,chen} which we
will show to contain the essential features necessary for addressing our
question. Section II discusses the equations and boundary conditions obeyed by
the modes of our 'universe'. The determination of the eigenfrequencies and
eigenfunctions requires the simultaneous diagonalization of a usually large
matrix and the solution of a transcendental equation. A numerical procedure
for a photonic lattice of arbitrary size and interface transmission is
presented in Appendix A.

In Sec. III, the radiation field is expanded in term of these large number of
modes and quantized. The radiation source is also treated quantum
mechanically, as an inhomogeneously broadened ensemble of two-level atoms
confined within the photonic lattice structure. The equations of motion for
the photon number and atomic populations are derived in this section. We
choose a fully quantized (i.e., quantized matter and field) treatment based on
Einstein's derivation of the Planck radiation law, which showed the importance
of a consistent treatment of stimulated and spontaneous emission processes.
\cite{einstein} Einstein was able to circumvent a fully quantized theory by
using Wien displacement law, which applied only to emission in free space. For
the photonic lattice, such a general relation does not exist.

There are several recent calculations of photonic-lattice emission where the
emitting source is a classically described current. \cite{luo,chen} An
advantage of our treatment over these classical ones is that by paralleling
the Planck radiation law derivation, our comparison of photonic-lattice and
blackbody emission spectra is appreciably more straightforward. \ Moreover,
the fluctuation-dissipation theorem \cite{louisell} which is an essential
assumption in the classical calculations, \cite{luo,chen} appears as a result
in a fully quantized treatment because spontaneous emission is treated from
first principles.

Section IV describes a spectrometer model used to determine the emission
spectrum. Section V uses the theory developed in the earlier sections to
investigate emission from an active photonic-lattice that is excited by an
external pump and allowed to equilibrate with a thermal bath via collisions.
The radiation field spectra measured inside and outside the photonic lattice
are described. Comparison of photonic lattice and blackbody emissions is
discussed for equilibrium and nonequilibrium situations.

Section VI summarizes the extension to a 3-dimensional geometry. \ The
comparison between photonic lattice and blackbody spectra is made assuming a
spherically symmetric photonic-lattice dielectric function. The 3-d treatment
is important for three reasons. First, it proves that the theory can retrieve
Planck's blackbody distribution in the absence of a photonic lattice. Second,
it verifies the 1-d treatment in terms of containing the necessary physics for
answering the question of photonic lattice versus blackbody thermal emission.
Lastly, it points out the substantial increase in numerical demands with
increase dimensionality, thus justifying our concentration on the 1-d analysis
to facilitate physical understanding and tractability of numerics.

\section{\bigskip Modes of the combined photonic-lattice and free-space
system}

In this section, the eigenmodes for a photonic lattice coupled to the outside
world are derived using the model depicted in Fig. 1. The universe, which
embeds the photonic lattice, is represented by a very large cavity with
perfectly reflecting walls at $z=0$ and $z=L$. (End results are extrapolated
by taking the limit $L\rightarrow\infty$.) The photonic lattice is modeled as
a series of coupled resonators with semitransparent interfaces. Following
Spencer and Lamb, \cite{spencer} the semitransparent interfaces are described
as very thin surfaces with very large dielectric constants. As an
idealization, we use dielectric 'bumps' giving a dielectric permittivity%
\begin{equation}
\epsilon\left(  z\right)  =\epsilon_{0}\left[  1+\frac{\eta}{\overline{k}}%
\sum\limits_{j=1}^{N_{pl}}\delta\left(  z-z_{j}\right)  \right]  \label{1}%
\end{equation}
where $\eta=2\sqrt{\left(  1-T_{pl}\right)  /T_{pl}}$, $T_{pl}$ is an
effective transmission at each interface located at $z_{j}$, $\overline{k}$ is
the average magnitude of the electromagnetic field wave vector and $N_{pl}$ is
the number of periods making up the photonic lattice. For brevity, we assume
the background permittivity inside the photonic lattice to be that of vacuum
$\epsilon_{0}$.

Using the above dielectric function in Maxwell equations gives the following
differential equation for the eigenmodes of the combined photonic-lattice and
free-space system:%
\begin{equation}
\frac{d^{2}}{dz^{2}}u_{k}\left(  z\right)  =-\mu_{0}\epsilon\left(  z\right)
\Omega_{k}^{2}u_{k}\left(  z\right)  \label{2}%
\end{equation}
where $\mu_{0}$ is the permeability in vacuum, $\Omega_{k}$ is the
eigenfrequency and $k$ labels the eigenmode. The boundary conditions are
obtained by first noting that the system is bounded by totally reflecting
surfaces, so that%
\begin{equation}
u_{k}(0)=u_{k}(L)=0 \label{3a}%
\end{equation}
Integrating Maxwell's equations across the bump gives the boundary conditions,%
\begin{align}
u_{k}(z_{j}^{+})  &  =u_{k}(z_{j}^{-})\label{3b}\\
\frac{d}{dz}u_{k}\left(  z_{j}^{+}\right)  -\frac{d}{dz}u_{k}\left(  z_{j}%
^{-}\right)   &  =-\eta ku_{k}(z_{j})\ , \label{3c}%
\end{align}
where the superscripts $-$ and $+$ indicate the positions immediately before
and after an interface, respectively. Integrating by parts (\ref{2}) gives the
orthogonality relation%
\begin{equation}
\int_{0}^{L}dz\ \epsilon\left(  z\right)  u_{k}\left(  z\right)  u_{l}\left(
z\right)  =\epsilon_{0}\delta_{k,l} \label{4}%
\end{equation}

Plotted in Fig. 2 are examples of eigenfunctions for a six-period ($N_{pl}=6$)
photonic lattice with effective interface transmission $T_{pl}=0.1$. Most of
the solutions are not resonant with the photonic lattice, so that mode
amplitude is negligible inside the photonic lattice, as shown in Fig. 2 (a).
Figures 2 (b) and 2 (c) show examples of photonic lattice modes, where the
latter figure clearly depicts the first derivative discontinuities at the interfaces.

We show in Figs. 3 and 4 that the model can reproduce the photonic-lattice
properties relevant to our study. Figure 3 illustrates the formation of bands
and bandgaps, by plotting the frequencies of the photonic lattice modes [i.e.,
modes depicted Figs. 2 (b) and 2 (c)] versus the interface transmission. Not
plotted are the large number of free-space modes [Fig. 2 (a)], with mode
separation $\Omega=m\pi c/L\longrightarrow0$ as the system length
$L\longrightarrow\infty$. At $T_{pl}=0$, the photonic-lattice modes are simply
the modes of six uncoupled resonators, each of length $a$, i.e., they are
$N_{pl}$-fold degenerate and have frequencies $\Omega=m\pi c/a$ where $m$ is
an integer. The degeneracy is removed with coupling among sections of the
photonic lattice. The result is groupings of states separated by energy gaps,
as shown in the figure. As $N_{pl}$ become very large, the groups of states
become continuous bands, with the photonic-lattice modes residing entirely
within the shaded regions, and the free-space modes residing outside.\ At
$T_{pl}=1,$ the model (with a very long $L$) approximates the free-space situation.

Figure 4 shows that the model can also reproduce the significant flattening of
the photonic-lattice dispersion at the bandedges. \ Plotted in the figure is
the dispersion for a twelve period ($N_{pl}=12$) photonic lattice, where the
points indicate the actual eigenfrequencies and the solid curve is a fit of
the data to illustrate the case of $N_{pl}\rightarrow\infty$. \ The flattening
of the dispersion at a bandedge results in a drastic increase in the density
of states. We define the density of states as $\rho(\omega)=dk_{0}/d\Omega$,
where following solid state convention $k_{0}$ is the wavevector with
vanishing interface reflectivity.\bigskip\ The effects of the large density of
states increase on intensity inside and outside of a photonic lattice is the
focus of this paper.

\section{\bigskip Active medium and radiation field}

To study the modification of emission characteristics by a photonic lattice,
we consider the situation of an ensemble of two-level atoms located inside a
photonic lattice. Each atom is labeled by $n$ and $j$, so that $\left\vert
a_{nj}\right\rangle $ and $\left\vert b_{nj}\right\rangle $ are the ground and
excited states, respectively, of an atom located at $z_{j}$ inside the
photonic lattice, with resonant energy $\hbar\omega_{n}$. Assuming the dipole
approximation, $z_{j}$ is a parameter locating the atom to a region that is
small compared to a wavelength, but large compared to the size of an atom. We
describe the radiation field emitted by these atoms in terms of the combined
system eigenmodes derived in the previous section, i.e.%

\begin{equation}
E\left(  z,t\right)  =\sum_{k}\mathcal{E}_{k}\left[  a_{k}\left(  t\right)
+a_{k}^{\dagger}\left(  t\right)  \right]  \ u_{k}\left(  z\right)  \label{5}%
\end{equation}
where $\mathcal{E}_{k}=\sqrt{\hbar\Omega_{k}/\left(  A\epsilon_{0}\right)  }$,
$a_{k}^{\dagger}$ and $a_{k}$ are the photon creation and annihilation
operators, respectively, and $A$ is the cross section area of the structure.
From (\ref{5}), using Maxwell's equations and a dipole interaction, the
Hamiltonian for the matter and radiation-field system is \cite{Quantum
Optics,john1}
\begin{equation}
H=\sum_{n,j}\hbar\omega_{n}\left\vert b_{nj}\right\rangle \left\langle
b_{nj}\right\vert +\sum_{k}\hbar\Omega_{k}a_{k}^{\dagger}a_{k}-\sum
_{k,n,j}g_{kj}\left(  \left\vert b_{nj}\right\rangle \left\langle
a_{nj}\right\vert a_{k}+a_{k}^{\dagger}\left\vert a_{nj}\right\rangle
\left\langle b_{nj}\right\vert \right)  \ , \label{6}%
\end{equation}
where $g_{kj}=\mu\mathcal{E}_{k}u_{k}\left(  z_{j}\right)  $ and $\mu$ is the
dipole matrix element. Introducing the operators for the microscopic
polarization amplitude $p_{njk}\equiv\left\vert b_{nj}\right\rangle
\left\langle a_{nj}\right\vert a_{k}\exp\left[  -i\left(  \omega_{n}%
-\Omega_{k}\right)  t\right]  $, the excited and ground state populations,
$\sigma_{anj}\equiv\left\vert a_{nj}\right\rangle \left\langle a_{nj}%
\right\vert $ and $\sigma_{bnj}\equiv\left\vert b_{nj}\right\rangle
\left\langle b_{nj}\right\vert $, respectively, and working in the Heisenberg
picture, \cite{louisell} we derive the equations of motion%

\begin{align}
\frac{dp_{njk}}{dt}  &  =\frac{i}{\hbar}e^{-i\left(  \omega_{n}-\Omega
_{k}\right)  t}\sum_{k^{\prime}}g_{k^{\prime}j}\left(  \sigma_{bnj}%
a_{k}a_{k^{\prime}}^{\dagger}-a_{k^{\prime}}^{\dagger}a_{k}\sigma
_{anj}\right)  \ \label{7}\\
\frac{d\sigma_{anj}}{dt}  &  =\frac{i}{\hbar}\sum_{k}g_{kj}\left[
p_{njk}^{\dag}e^{-i\left(  \omega_{n}-\Omega_{k}\right)  t}-p_{njk}e^{i\left(
\omega_{n}-\Omega_{k}\right)  t}\right] \label{8}\\
\frac{d\sigma_{bnj}}{dt}  &  =-\frac{i}{\hbar}\sum_{k}g_{kj}\left[
p_{njk}^{\dag}e^{-i\left(  \omega_{j}-\Omega_{k}\right)  t}-p_{njk}e^{i\left(
\omega_{j}-\Omega_{k}\right)  t}\right]  \label{9}%
\end{align}
Additionally, the photon number operator obeys,
\begin{equation}
\frac{da_{k}^{\dagger}a_{k}}{dt}=\frac{i}{\hbar}\sum_{n,j}g_{kj}\left[
p_{njk}^{\dag}e^{-i\left(  \omega_{j}-\Omega_{k}\right)  t}-p_{njk}e^{i\left(
\omega_{j}-\Omega_{k}\right)  t}\right]  \ . \label{10}%
\end{equation}
Assuming that the polarization decays because of dephasing collisions and that
the effective decay rate $\gamma$ is much larger than the rate of changes in
the active medium and photon populations, we can adiabatically eliminate the
polarization equation. Then, introducing the expectation values%

\begin{align}
N_{k}  &  =\left\langle a_{k}^{\dagger}a_{k}\right\rangle \label{11}\\
N_{an}  &  =\sum_{j=1}^{N}\left\langle \sigma_{anj}\right\rangle \label{12}\\
N_{bn}  &  =\sum_{j=1}^{N}\left\langle \sigma_{bnj}\right\rangle \label{13}%
\end{align}
we obtain the working equations for our analysis:%

\begin{align}
\frac{dN_{an}}{dt}  &  =\frac{2\mu^{2}}{\hbar\epsilon_{0}AL_{c}\gamma}%
\sum\limits_{k}\Omega_{k}\ \Gamma_{k}\ \left[  \left(  N_{bn}-N_{an}\right)
N_{k}+N_{bn}\right]  \ L\left(  \omega_{n}-\Omega_{k}\right) \nonumber\\
&  -\gamma_{r}\left[  N_{an}-f_{a}\left(  \omega_{n},T\right)  \right]
-\Lambda\left(  \omega_{n}\right)  N_{an}\label{14}\\
\frac{dN_{bn}}{dt}  &  =-\frac{2\mu^{2}}{\hbar\epsilon_{0}AL_{c}\gamma}%
\sum\limits_{k}\Omega_{k}\ \Gamma_{k}\ \left[  \left(  N_{bn}-N_{an}\right)
N_{k}+N_{bn}\right]  \ L\left(  \omega_{n}-\Omega_{k}\right) \nonumber\\
&  -\gamma_{r}\left[  N_{bn}-f_{b}\left(  \omega_{n},T\right)  \right]
+\Lambda\left(  \omega_{n}\right)  N_{an}\label{15}\\
\frac{dN_{k}}{dt}  &  =\frac{2\mu^{2}}{\hbar\epsilon_{0}AL_{c}\gamma}\sum
_{n}\Omega_{k}\ \Gamma_{k}\ \left[  \left(  N_{bn}-N_{an}\right)  N_{k}%
+N_{bn}\right]  L\left(  \omega_{n}-\Omega_{k}\right)  -\gamma_{c}N_{k}
\label{16}%
\end{align}
where $N$ is the number of atoms, $L\left(  x\right)  =\left[  1+\left(
x/\gamma\right)  ^{2}\right]  $ and%
\begin{equation}
\Gamma_{k}=\int_{0}^{L_{c}}dz\ \left\vert u_{k}\left(  z\right)  \right\vert
^{2} \label{17}%
\end{equation}
is the mode confinement factor. In (\ref{14}) - (\ref{16}), the pump and decay
contributions are included phenomenologically, $\gamma_{c}$ is the photon
decay rate, $\Lambda\left(  \omega_{n}\right)  =\Lambda_{0}\exp\left[
\hbar\left(  \omega_{0}-\omega_{n}\right)  /k_{B}T_{p}\right]  $ is the pump
rate, $\hbar\omega_{0}$ is the material bandgap energy and $\gamma_{r}$ is an
effective rate for the actual populations $N_{an}$ and $N_{bn}$ to relax to
the equilibrium distributions%

\begin{align}
f_{a}(\omega_{n},T)  &  =Z_{n}\label{18}\\
\text{\ }f_{b}(\omega_{n},T)  &  =Z_{n}\exp%
\genfrac{(}{)}{}{}{-\hbar\omega_{n}}{k_{B}T}%
\ , \label{19}%
\end{align}
where%
\begin{equation}
Z_{n}=\left[  1+\exp%
\genfrac{(}{)}{}{}{-\hbar\omega_{n}}{k_{B}T}%
\right]  ^{-1}\ , \label{21}%
\end{equation}
$T_{p}$ and $T$ \ are the pump and reservoir temperatures. \ In our study,
(\ref{14}) to (\ref{16}) are solved numerically.

\section{\bigskip Detector}

To determine the spectra of the intracavity and output radiation, we use the
simple spectrometer model shown in Figure 5. In this model, two-level atoms
are placed in the region of interest. These atoms are prepared with only the
ground state $\left\vert a_{n}^{d}\right\rangle $ populated when the radiation
field is absent (zero detector temperature). The label $n$ indicates that the
level spacing between $\left\vert b_{n}^{d}\right\rangle $ and $\left\vert
a_{n}^{d}\right\rangle $ is $\omega_{n}^{d}$. The atoms interact weakly with
the radiation field to be measured, which excites some fraction of the atoms
to an excited state $\left\vert b_{n}^{d}\right\rangle $ that has some finite
lifetime $\gamma_{d}^{-1}$. Assuming a sufficiently fast detector response so
that the detector populations adiabatically follow the variations in the
photon number, the population in state $\left\vert b_{n}^{d}\right\rangle $
gives a measure of the radiation intensity ($\propto N_{k}$) in the region
occupied by the detector atom. The steady state upper detector state
population is%
\begin{equation}
N_{b}^{d}\left(  \omega_{n}^{d}\right)  =D\sum\limits_{k}\Omega_{k}N_{k}%
\frac{\gamma_{d}}{\gamma_{d}^{2}+\left(  \omega_{n}^{d}-\Omega_{k}\right)
^{2}}\int_{z_{d}}^{z_{d}+L_{d}}dz\ \left\vert u_{k}\left(  z\right)
\right\vert ^{2} \label{22}%
\end{equation}
where $D=2\mu_{d}N_{d}/\left(  \hbar\epsilon_{0}AL_{d}\gamma_{d}\right)  $,
$\mu_{d}$ is the dipole matrix element between states $\left\vert b_{n}%
^{d}\right\rangle $ and $\left\vert a_{n}^{d}\right\rangle $, $L_{d}$ is the
length of the detected region, and $N_{d}$ is the number of detector atoms.
Measuring this population for atoms of different $\omega_{n}^{d}$ gives the
spectrum within the region $z_{d}\leq z\leq z_{d}+L_{d}$. In this model,
$N_{d}$ and the decay rate $\gamma_{b}$ should be sufficiently large to
prevent saturation of the detector. \ On the other hand, too large a
$\gamma_{d}$ degrades spectral resolution. Alternately, one may use two level
atoms injected into the region of interest, and removed after a short time.
\cite{scully2}

\section{Photonic lattice emission}

We consider a twelve-period photonic lattice with $L_{c}=120\mu m$, $L=1.2cm$
and interface transmission $T_{pl}=0.01,$ $0.1$ and $0.4$. The eigenmodes are
determined by solving (\ref{2}) with the boundary conditions (\ref{3a}) -
(\ref{3c}). The results are used in (\ref{14}) - (\ref{16}), which are solved
numerically with a fourth-order Runge-Kutta finite difference method. The
input parameters are $\gamma=10^{12}s^{-1}$, $\gamma_{c}=10^{9}s^{-1}$,
$\Lambda_{0}=10^{10}s^{-1}$, $\omega_{0}=1.6\times10^{14}s^{-1}$, $\mu
=e\times1.3nm$, $N=601$ and $T_{p}=T=400K$.

\subsection{Equilibrium}

To relate to earlier studies, \cite{dowling,narayanaswamy,trupke,luo} we first
compare photonic lattice and blackbody emissions under thermal equilibrium
conditions. To do so, we perform the calculations for a rapid population
relaxation rate of $\gamma_{r}=10^{13}s^{-1}$, which ensures (verified after
the time integration) that the steady-state active-medium populations $N_{an}$
and $N_{bn}$ are to a good approximation given by the equilibrium
distributions $f_{a}(\omega_{n},T)$ and $f_{b}(\omega_{n},T)$, respectively.
The solid curves in Fig. 6 (a) show the calculated intracavity emission
spectra for three interface transmissions. \ In the figure, we define an
intracavity detector signal,%
\begin{equation}
S_{in}\left(  \omega\right)  \equiv\frac{N_{b}^{d}\left(  \omega\right)  }%
{D}=\sum\limits_{k}\Omega_{k}N_{k}\frac{\gamma_{d}}{\gamma_{d}^{2}+\left(
\omega-\Omega_{k}\right)  ^{2}}\int_{0}^{L_{c}}dz\ \left\vert u_{k}\left(
z\right)  \right\vert ^{2} \label{23}%
\end{equation}
where $N_{b}^{d}\left(  \omega\right)  $ is calculated using the steady-state
solution for $N_{k}$ in (\ref{22}). The figure shows two bands of
photonic-lattice states, where the frequency extent of the bands depends on
the interface transmission. Between the two bands is a photonic bandgap where
emission is strongly suppressed. By repeating the calculation with $T_{pl}=1$,
we obtain the corresponding blackbody spectrum (dashed curve). Comparison of
the curves clearly indicates the significant intensity enhancement inside a
photonic lattice, especially at the bandedges for $T_{pl}=0.01$.

To determine the output spectrum, we place the spectrometer in the free-space
region. Figure 7 shows the output detector signal,
\begin{equation}
S_{out}\left(  \omega\right)  \equiv\frac{N_{b}^{d}\left(  \omega\right)  }%
{D}=\sum\limits_{k}\Omega_{k}N_{k}\frac{\gamma_{d}}{\gamma_{d}^{2}+\left(
\omega-\Omega_{k}\right)  ^{2}}\int_{L-L_{c}}^{L}dz\ \left\vert u_{k}\left(
z\right)  \right\vert ^{2} \label{24}%
\end{equation}
for the same interface transmissions as in Fig. 6. In contrast to inside the
photonic lattice, where there is significant optical intensity enhancement,
Fig. 7 indicates that the intracavity emission peaks are appreciably depressed
outside the photonic lattice. The strong intracavity enhancement by the
photonic-lattice density of states appears to be cancelled by an outcoupling
attenuation. This leaves the photonic lattice emission peaks to be essentially
independent of interface transmission. More importantly, these peaks lie at or
slightly below the blackbody emission curve.

Simulations performed over a wide range of input parameters point towards the
result that as long as the active-medium populations $N_{an}$ and $N_{bn}$ are
in thermal equilibrium, the photonic-lattice output is always below that of
the blackbody. For instance, the spectra are insensitive to the choice of
$\gamma$. The blackbody spectrum is also insensitive to $\gamma_{d}$ because
of the weak frequency dependence of the blackbody photon density. However, for
the photonic-lattice, too large a $\gamma_{d}$ degrades the spectrometer
resolution and leads to lower spectral peaks. For the opposite situation, too
small a $\gamma_{d}$ introduces noise in the spectrum because of the
inadequate resolution of the system normal modes (i.e. because $L$ is
insufficiently large).

\subsection{Nonequilibrium}

To study active photonic-lattice operation in greater generality, we allow the
active medium populations to deviate from thermal equilibrium. The
investigation is performed by repeating the earlier calculations, keeping all
input parameters except $\gamma_{r}$ the same. Figure 8 illustrates the
changes in the excited-state population distribution $N_{bn}$, at steady state
and for decreasing population relaxation. When $\gamma_{r}$ is reduced to
$10^{11}s^{-1}$ from $10^{13}s^{-1}$, a slight difference emerges between the
excited-state populations of the identically pumped photonic-lattice and
blackbody active media. The solid curve in Fig. 8 (a) shows a noticeable
deformation of the photonic-lattice excited-state population distribution.
There is also a significant difference between $N_{bn}$ and $f_{b}\left(
\omega_{n},T\right)  $ for $T=400K$ (dot-dashed curve), which are the actual
distribution and the asymptotic ($\gamma_{r}\rightarrow\infty$) equilibrium
distribution, respectively. Further reduction to $\gamma_{r}=10^{10}s^{-1}$
significantly increases the deviation of the photonic-lattice excited-state
population distribution from a Maxwell Boltzmann distribution [see solid
curve, Fig 8 (b)]. Holes are burned in the distribution because the population
relaxation is insufficiently fast to replenish the excited-state population
depleted by the spectrally relatively narrow radiation field emitted by
photonic lattice. There is also a change in the blackbody distribution [dashed
curve, Fig. 8(b)], to one that approximates a Maxwell Boltzmann distribution
at $T\approx500K$ (dotted curve). Since the active media in both structures
are identical, the difference between the photonic-lattice and blackbody
populations (solid and dashed curves, respectively) is from photonic-lattice effects.

The effects of the population changes in Fig. 8 on the emission spectra are
depicted in Fig 9. Plotted on the y-axis is the relative emission intensity
inside (outside) the photonic lattice, which we define as $S_{in(out)}\left(
\omega\right)  $ for the photonic lattice divided by $S\left(  \omega\right)
$ for the blackbody. In spite of the large increase in excited state
population, we find that the intracavity and output relative intensities
remain basically unchanged when $\gamma_{r}$ is reduced from $10^{13}s^{-1}$
to $10^{11}s^{-1}$. In particular, the output photonic-lattice intensity
remains at or slightly below that of the blackbody (i.e., relative intensity
$\leq1$). However, the result changes considerably for $\gamma_{r}%
=10^{10}s^{-1}$. Here, the intensity within the photonic-lattice band
increases considerably relative to that of the blackbody both inside and
outside the cavity (solid curves). More importantly, the solid curve in Fig. 9
(b) clearly shows greater output intensity for the photonic lattice than the
blackbody throughout the emission band of the photonic lattice. This
enhancement of output emission occurs for identically pumped active regions
and is a result of a nonequilibrium population that shows significant hole
burning. The presence of nonequilibrium effects may be the cause for
experimental observations of metallic photonic-lattice emission exceeding that
of the blackbody. \cite{lin3} Note that the difference in output emission
spectra [solid and dashed curves in Fig. 9 (b)] comes from population
distributions that are, on the average, quite similar. That is, a
least-squares fit of the solid and dashed curves in Fig. 8 (b) will produce
Maxwell-Boltzmann's distributions that differ in temperature by less than 20K.
Therefore, measurement of average temperature will not identify the
experimental conditions leading to the photonic-lattice output emission
exceeding that of the blackbody. Rather, an energy-resolved measurement of the
emitter upper state population is necessary.

\ The end results reached in our analyses involving equilibrium and
nonequilibrium situations are robust, i.e., they are relatively insensitive to
the choice of input parameters. While calculations performed with different
interface transmission show significant differences in spectral shapes and
intracavity intensities, the output intensities remain relatively constant
because of the mitigating influence of the coupling to free space.
Calculations are also performed for different number of photonic-lattice
periods. The results show negligible differences beyond $N_{pl}=10$, thus
verifying that the use of a 12-period photonic lattice does not lead to loss
of generality. Clearly noticeable are effects, such as differences in excess
output intensity with varying interface transmission, that are due to optical
nonlinearities in the nonequilibrium active medium. \ Such effects will not be
present in treatments using linear classical sources. \cite{luo} 

\section{ Extension to a 3-dimensional photonic lattice{}}

\bigskip This section treats a 3-dimensional photonic lattice. Following
earlier quantum optical studies of photonics lattices, \cite{john} a
spherically symmetric dielectric function is assumed to simplify the numerics.
In spherical coordinates, the equation satisfied by the passive eigenmodes of
the combined photonic-lattice and free-space system, $u_{klm}\left(
r,\theta,\phi\right)  $, is%
\begin{equation}
\frac{1}{r}\frac{\partial^{2}}{\partial r^{2}}\left(  ru_{klm}\right)
+\frac{1}{r^{2}\sin\theta}\frac{\partial}{\partial\theta}\left(  \sin
\theta\frac{\partial u_{klm}}{\partial\theta}\right)  +\frac{1}{r^{2}\sin
^{2}\theta}\frac{\partial^{2}u_{klm}}{\partial\phi^{2}}=-\mu_{0}%
\epsilon\left(  r\right)  \Omega_{klm}^{2}u_{klm}\ . \label{25}%
\end{equation}
Choosing the dielectric function%
\begin{equation}
\epsilon\left(  r\right)  =\epsilon_{0}\left[  1+\frac{\eta}{\overline{k}}%
\sum\limits_{j=1}^{N_{pl}}\delta\left(  r-r_{j}\right)  \right]  \ ,
\label{26}%
\end{equation}
\bigskip where $\eta$ and $\overline{k}$ are the same as in (\ref{1}), a
solution of (\ref{25}) between the photonic-lattice interfaces or in the
free-space region is%

\begin{equation}
u_{klm}\left(  r,\theta,\phi\right)  =\left[  A_{kln}j_{l}\left(  kr\right)
+B_{kln}\eta_{l}\left(  kr\right)  \right]  Y_{lm}\left(  \theta,\phi\right)
\ , \label{27}%
\end{equation}
where $j_{l}\left(  \rho\right)  $ and $\eta_{l}\left(  \rho\right)  $ are
spherical Bessel and Neumann functions, $Y_{lm}\left(  \theta,\phi\right)  $
is a spherical harmonic and the subscript $n$ indicates that the coefficients
$A_{kln}$ and $B_{kln}$ are for $r_{n}<r\leq r_{n+1}$. In order for a solution
to be finite at the origin and vanish at $r=r_{N_{pl}+1}$ (the end of the
region representing free space), we require
\begin{equation}
B_{kl1}=0 \label{28}%
\end{equation}

\begin{equation}
A_{klN_{pl}+1}j_{l}\left(  kr\right)  +B_{klN_{pl}+1}\eta_{l}\left(
kr\right)  =0\ . \label{29}%
\end{equation}
At the interfaces, the boundary conditions (\ref{3b}) and (\ref{3c}) demand%
\begin{equation}
A_{kln}j_{l}\left(  kr\right)  +B_{kln}\eta_{l}\left(  kr\right)
-A_{kln+1}j_{l}\left(  kr\right)  -B_{kln+1}\eta_{l}\left(  kr\right)  =0
\label{30}%
\end{equation}%
\begin{gather}
A_{kln}\left[  -j_{l}^{\prime}\left(  kr\right)  +\eta j_{l}\left(  kr\right)
\right]  +B_{kln}\left[  -\eta_{l}^{\prime}\left(  kr\right)  +\eta\eta
_{l}\left(  kr\right)  \right] \nonumber\\
+A_{kln+1}j_{l}^{\prime}\left(  kr\right)  +B_{kln+1}\eta_{l}^{\prime}\left(
kr\right)  =0 \label{31}%
\end{gather}
for $2\leq n\leq N_{pl}$ and $r=r_{1},r_{2,...,}r_{N_{pl}}$.

The numerical solution is implemented similar to what is described in Appendix
A. A 6-period photonic lattice is considered, where the lattice constant is
$2\mu m$ and the interface transmission is $T_{pl}=0.05$. Coupled to the
photonic lattice is a 'free'-space region extending from $24\mu m<r\leq612\mu
m$. System dynamics is governed by the equations of motion (\ref{14}) -
(\ref{16}) with the photon-state index $k$ replaced by the three indices,
$k,l$ and $m$. Numerical analyses of the steady state solutions are performed
assuming the input parameters, $\gamma=2\times10^{12}s^{-1}$, $\gamma
_{r}=10^{13}s^{-1}$, $\gamma_{c}=10^{12}s^{-1}$, $\Lambda_{0}=10^{12}s^{-1}$,
$\omega_{0}=1.6\times10^{14}s^{-1}$, $\mu=e\times1.3nm$, $N=601$ and
$T_{p}=T=200K.$ Similar to the 1-dimensional case, the rates are chosen to
ensure reaching steady state with active-medium populations $N_{an}$ and
$N_{bn}$ described by Maxwell-Boltmann distributions.

To obtain the intracavity emission spectrum, detector atoms are placed inside
the photonic-lattice structure. The probability of finding a photon with
frequency $\omega$ is proportional to%
\begin{equation}
S_{in}\left(  \omega\right)  =\sum_{klm}\Omega_{klm}N_{klm}\frac{\gamma_{d}%
}{\gamma_{d}^{2}+\left(  \omega-\Omega_{klm}\right)  ^{2}}\int_{0}^{2\pi}%
d\phi\int_{0}^{\pi}d\theta\int_{0}^{r_{Npl}}dr\ r^{2}\ \left\vert
u_{klm}\left(  r,\theta,\phi\right)  \right\vert ^{2}\ ,\label{32a}%
\end{equation}
so that the emission energy is proportional to $\hbar\omega S_{in}\left(
\omega\right)  $. \ In Fig. 10, the solid curve is a plot of $\omega S_{in}$
as a function of frequency. It shows four narrow emission bands separated by
photonic bandgaps. Repeating the calculation using interface transmission
$T_{pl}=1$, gives the free-space emission spectrum (dashed curve). Examination
of the populations after steady state is reached verifies that both solid and
dashed curves are for identical Maxwell-Boltmann distributions at $T=200K$.
The curves clearly indicate intensity enhancement inside the photonic lattice.

\bigskip To compare the output emission in a given direction, the detector
atoms are placed to give a signal,
\[
S_{out}\left(  \omega\right)  =\sum_{klm}\Omega_{klm}N_{klm}\frac{\gamma_{d}%
}{\gamma_{d}^{2}+\left(  \omega-\Omega_{klm}\right)  ^{2}}\int_{\phi_{d1}%
}^{\phi_{d2}}d\phi\int_{\theta_{_{d1}}}^{\theta_{_{d2}}}d\theta\int_{r_{d1}%
}^{r_{d2}}dr\ r^{2}\ \left\vert u_{klm}\left(  r,\theta,\phi\right)
\right\vert ^{2}\ ,
\]
where $r_{d1}$, $r_{d2}$ are within the free-space region, $\theta_{d1}%
,\theta_{d2}$ and $\phi_{d1},\phi_{d2}$ define the direction and collection
solid angle. The solid curve and dashed curves in Fig. 11 shows the output
photon-lattice and free-space emission spectra, respectively. These results
are obtained for $r_{d1}=100\mu m$, $r_{d2}=124\mu m$, $\phi_{d1}=\theta
_{d1}=0,$ $\phi_{d2}=2\pi$ and $\theta_{d2}=\pi/18$, which define emission
within a cone of $\pm10^{\circ}$ in the z-direction. Comparison of solid and
dashed curves indicates that, similar to the 1-dimensional case, peak
intensities measured outside the photonic-lattice structure do not exceed
those of the blackbody.

\bigskip A test of our treatment is to see how well it reproduces Planck's
distribution for the frequency spectrum of free-space emission energy from a
thermal source. \ The dotted curve in Fig. 11 is proportional to $\omega
^{3}\left[  \exp\left(  \hbar\omega/k_{B}T\right)  -1\right]  ^{-1}$ with
temperature $T=200K$, and it depicts the shape of the blackbody frequency
spectrum according to Planck's formula. The agreement is good, considering
that there are several factors causing discrepancies. Two important ones are
truncating the optical modes\ at $l=13$ and limiting 'free'-space to $24\mu
m<r\leq612\mu m$, in order to maintain reasonable computation times. \ Even
so, over $10^{4}$ optical modes are used. Other factors contributing to the
differences include the presence of optical loss [$\gamma_{c}\neq0$ in
(\ref{16})], which is neglected in the derivation of Planck's distribution.
While increasing the number of optical modes improves agreement with Planck's
formula, it does not impact the blackbody versus photonic lattice emission
comparison. This is because the result of the photonic-lattice output
intensity spectrum being bounded inside the blackbody one applies separately
for each $l$.

\section{ VI. Conclusion}

In summary, the emission from an active photonic lattice is investigated using
a model consisting of an inhomogeneously broadened ensemble of two-level atoms
interacting with a multimode radiation field. A fully quantized (i.e.,
quantized atoms and quantized electromagnetic field) description is chosen to
provide a consistent description of stimulated and spontaneous emission.
Furthermore, to describe the modal properties of the radiation field of a
finite photonic lattice coupled to free space, the analysis considers the
photonic lattice and free space as one combined system. This circumvents a
long-standing inconsistency in quantum optics involving the decoupling of the
treatments of the cavity normal modes and outcoupling losses.

Our approach gives the emission spectra for arbitrary photonic-lattice
configurations and reproduces Planck's blackbody radiation formula for thermal
emission in free space. Comparison of photonic-lattice and blackbody emission
shows appreciable modification of the blackbody spectrum by the photonic
lattice, where the redistribution of the photon density of states results in
suppression of radiation at certain wavelengths and enhancement at others. The
enhancement can give rise to high intracavity intensity peaks, especially at
the photonic-lattice bandedges. These intensity peaks are mitigated outside
the photonic lattice by the spectrally dependent outcoupling. For population
relaxation sufficiently fast to ensure the same equilibrium population
distribution in both structures, the photonic-lattice output intensity does
not exceed that of the blackbody at the same frequency. However, for slow
population relaxation, there is a greater tendency for a nonequilibrium
photonic-lattice population. Then, in the presence of population hole burning,
the intensity in certain regions of the photonic-lattice spectrum can exceed
that of the blackbody, even when both structures are identically pumped.

\section{ V. Acknowledgments}

This work was supported by the U. S. Department of Energy under contract No.
DE-AC04-94AL85000 and by a Senior Scientist Award from the Alexander von
Humboldt Foundation. \ The author thanks I. El-Kady, I. Waldmueller and S.
Wieczorek for helpful discussions.

\section{ Appendix A: Numerical evaluation of system eigenmodes}

In this appendix we describe a numerical procedure for evaluating the
eigenmodes of the combined photonic-lattice and free-space system. This
procedure applies for a photonic lattice of arbitrary size and interface
transmission. In region $n$, which may be any section of the photonic lattice
or the section representing free space, the solutions of (\ref{2}) have the form%

\begin{equation}
u_{k}\left(  z_{n}\right)  =A_{k,n}\sin\left(  kz_{n}\right)  +B_{k,n}%
\cos\left(  kz_{n}\right)  \tag{A1}%
\end{equation}
where $k=\Omega_{k}/c$. Because of boundary condition (\ref{3a}),%
\begin{align}
A_{k,1}  &  \neq0\tag{A2}\\
B_{k,1}  &  =0 \tag{A3}%
\end{align}%
\begin{equation}
A_{k,N_{pl}+1}\sin\left(  kz_{N_{pl}+1}\right)  +B_{k,N_{pl}+1}\cos\left(
kz_{N_{pl}+1}\right)  =0 \tag{A4}%
\end{equation}
From boundary conditions (\ref{3b}) and (\ref{3c}),%
\begin{align}
A_{k,n}\sin\left(  kz_{n}\right)  +B_{k,n}\cos\left(  kz_{n}\right)   &
=A_{k,n+1}\sin\left(  kz_{n}\right)  +B_{k,n+1}\cos\left(  kz_{n}\right)
\tag{A5}\\
A_{k,n}\left[  -\cos\left(  kz_{n}\right)  +\eta\sin\left(  kz_{n}\right)
\right]  +B_{k,n}\left[  \sin\left(  kz_{n}\right)  +\eta\cos\left(
kz_{n}\right)  \right]   &  =-A_{k,n+1}\cos\left(  kz_{n}\right)
+B_{k,n+1}\sin\left(  kz_{n}\right)  \tag{A6}%
\end{align}
for $2\leq n\leq N_{pl}$.

There are many approaches to numerically solve the above equations. We
describe below the one we followed in this paper. \ Basically, we look for the
values of $k$ satisfying (A4), where the mode amplitudes $A_{k,N_{pl}+1}$ and
$B_{k,N_{pl}+1}$ are obtained by solving a $2N_{pl}\times2N_{pl}$ matrix
equation
\begin{equation}
SU=D \tag{A7}%
\end{equation}
The matrix elements of $S$ are best defined by separating the even and odd
number rows. For $n=odd$,
\begin{align}
S_{i,i}  &  =-\sin\left(  kz_{1}\right) \tag{A8}\\
S_{i,i+1}  &  =-\cos\left(  kz_{i}\right)  \tag{A9}%
\end{align}
and for $n=even$,
\begin{equation}
S_{i,i}=-\sin\left(  kz_{i-1}\right)  \tag{A10}%
\end{equation}
For $n\geq3$ and $n=odd$,%

\begin{align}
S_{i,i-1}  &  =\cos\left(  kz_{i}\right) \tag{A11}\\
S_{i,i-2}  &  =\sin\left(  kz_{i}\right)  \tag{A12}%
\end{align}
and for $n\geq4$ and $n=even$,%

\begin{align}
S_{i,i-1}  &  =\cos\left(  kz_{i-1}\right) \tag{A13}\\
S_{i,i-2}  &  =\sin\left(  kz_{i-1}\right)  +\eta\cos\left(  kz_{i-1}\right)
\tag{A13}\\
S_{i,i-3}  &  =-\cos\left(  kz_{i-1}\right)  +\eta\sin\left(  kz_{i-1}\right)
\tag{A14}%
\end{align}
All other matrix elements are zero. The elements of the column matrix $D$
vanishes except for%
\begin{align}
D_{1}  &  =-\sin\left(  kz_{1}\right) \tag{A15}\\
D_{2}  &  =\cos\left(  kz_{2}\right)  -\eta\sin\left(  kz_{2}\right)
\tag{A16}%
\end{align}

Equation (A7) is solved using the Gauss-Jordan method. In the solutions and
for $j=odd$, $U_{j}$ gives the coefficient $A_{k,j+1},$while $U_{j+1}$ gives
the coefficient $B_{k,j+1}$. At this stage, we have set $A_{k,1}=1$ so that
the eigenfunctions are unnormalized. \ We perform the normalization according
to (\ref{4}).

\pagebreak

\begin{center}
\bigskip\textbf{Figure Captions\\[2mm]}
\end{center}

Fig. 1. Model of a photonic lattice connected to a large cavity approximating
the universe.

Fig. 2. \ Eigenfunctions of a 6-period photonic-lattice coupled to free space
for interface transmission $T_{pl}=0.10$. The figure shows non-resonant (a)
and resonant (b and c) photonic-lattice modes.

Fig. 3. Eigenfrequency versus interface transmission for a 6-period photonic
lattice. The points indicate the actual eigenmodes of the finite structure,
while the shaded regions illustrate the extent of the bands of
photonic-lattice states when the number of periods become very large.

Fig. 4. Dispersion for a 12-period photonic lattice. The points are the
eigenmodes, the solid curve is a fit through these points and the dashed curve
shows the free-space dispersion.

Fig. 5. Spectrometer model where the upper level decay prevents detector
saturation and approximates the drift of carriers to the electrodes in a
reverse-baised photodiode.

Fig. 6. Photonic-lattice (solid curves) and blackbody (dashed curves)
intracavity emission spectra.

Fig. 7. Photonic-lattice (solid curves) and blackbody (dashed curves) output
emission spectra.

Fig. 8. Upper state population for photonic lattice (solid curve) and
blackbody (dashed curve) versus transition frequency for $\gamma_{r}=10^{11}$
(a) and $10^{10}s^{-1}$ (b). The dot-dashed curves show the equilibrium
distribution at $400K$. The dotted curve is the equilibrium distribution at
$500K$.

Fig. 9. Relative intensity spectra inside (a) and outside (b) the photonic
lattice in Fig. 8. The curves are for $\gamma_{r}=10^{11}$ (dashed) and
$10^{10}s^{-1}$ (solid). Above the long-dashed line, the photonic-lattice
intensity is higher than the blackbody's.

Fig. 10. 3-dimensional photonic-lattice and blackbody intracavity emission
spectra (solid and dashed curves, respectively).

Fig. 11. Output emission spectra for 3-d photonic-lattice (solid curve),
blackbody (dashed curve) and Planck's distribution (dotted curve).
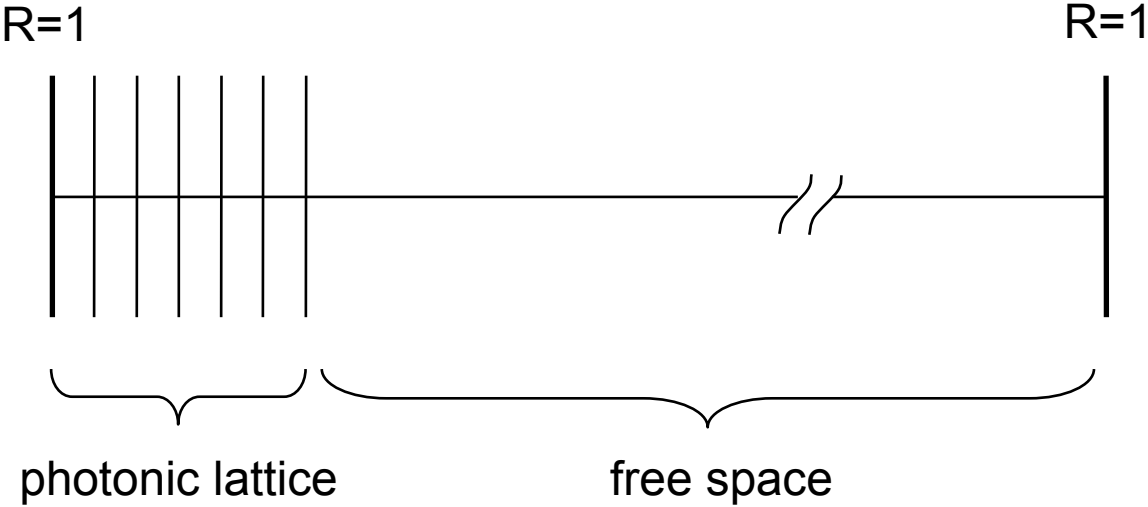
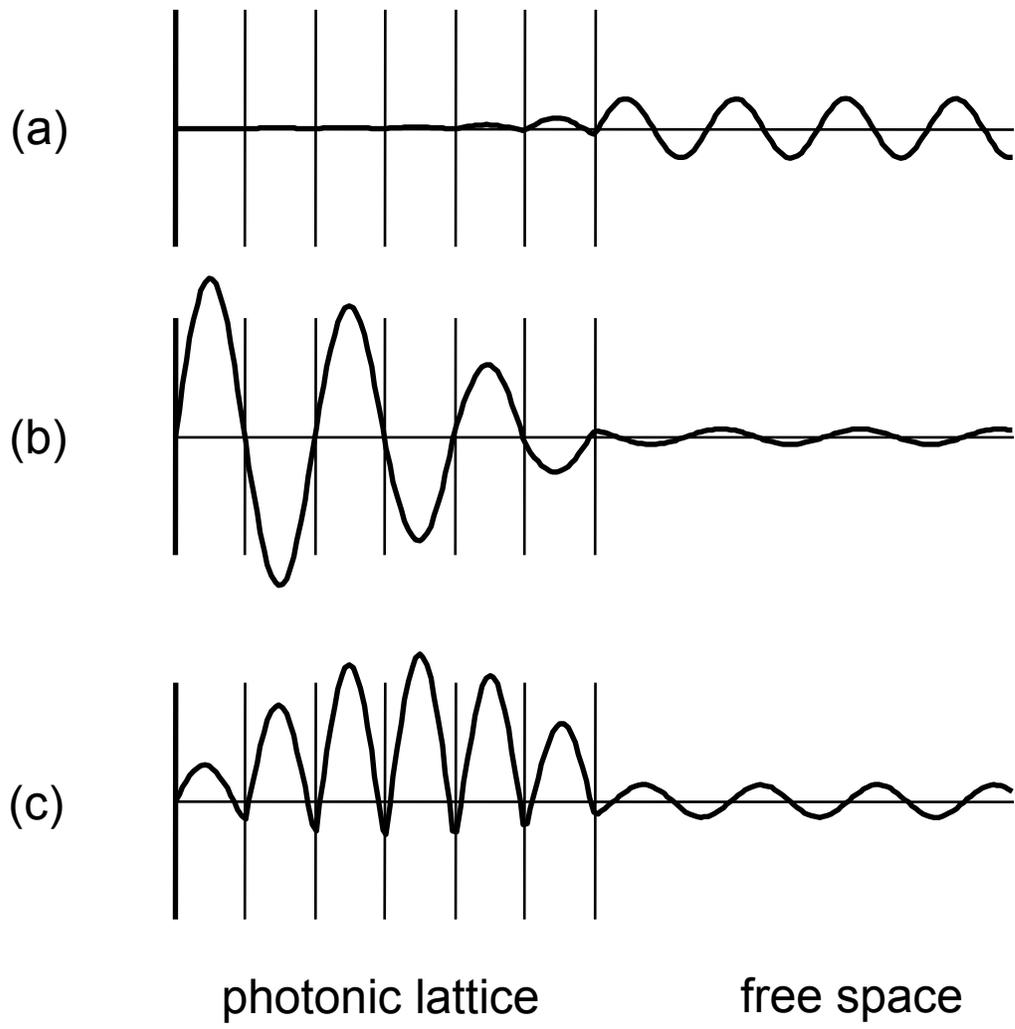
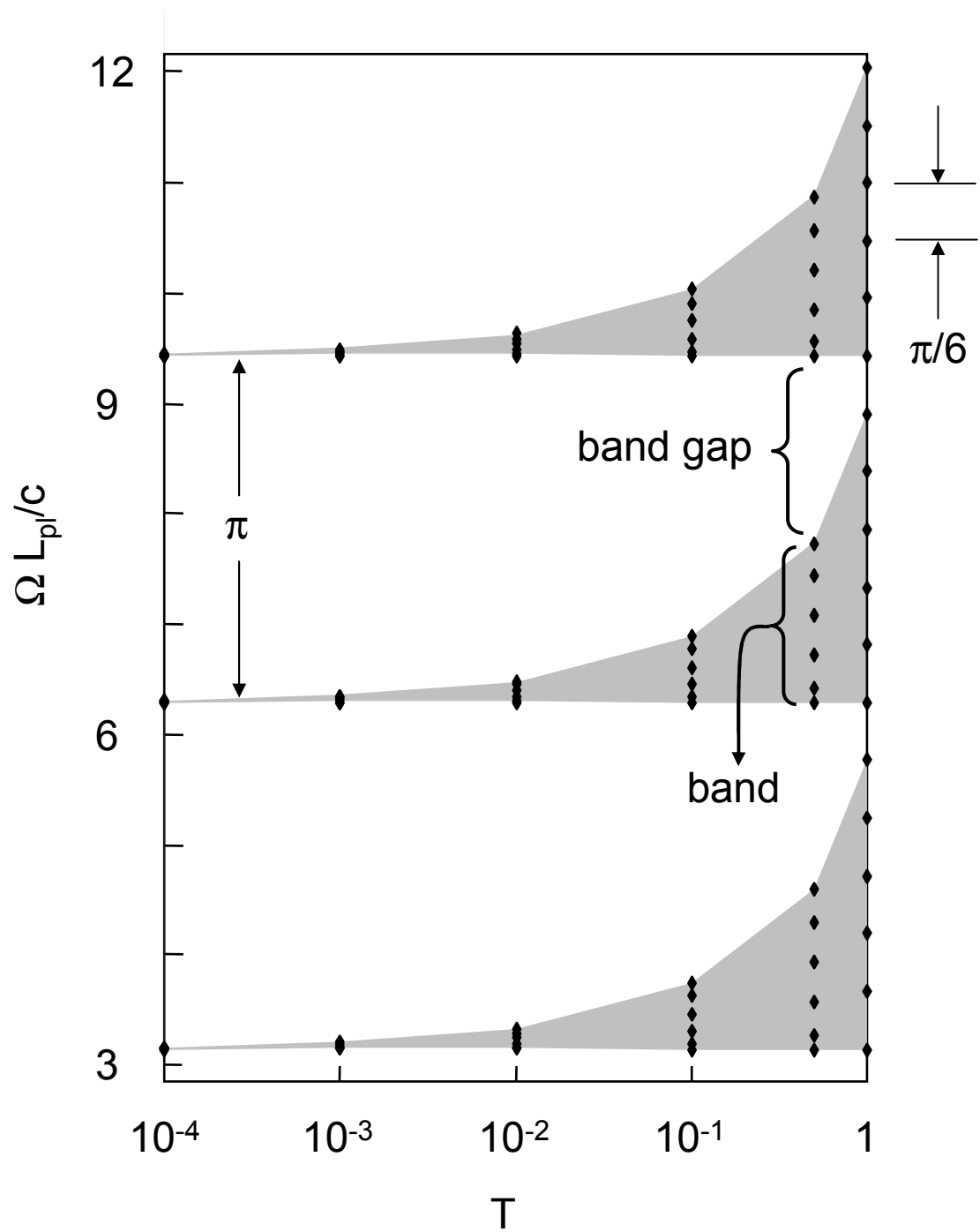
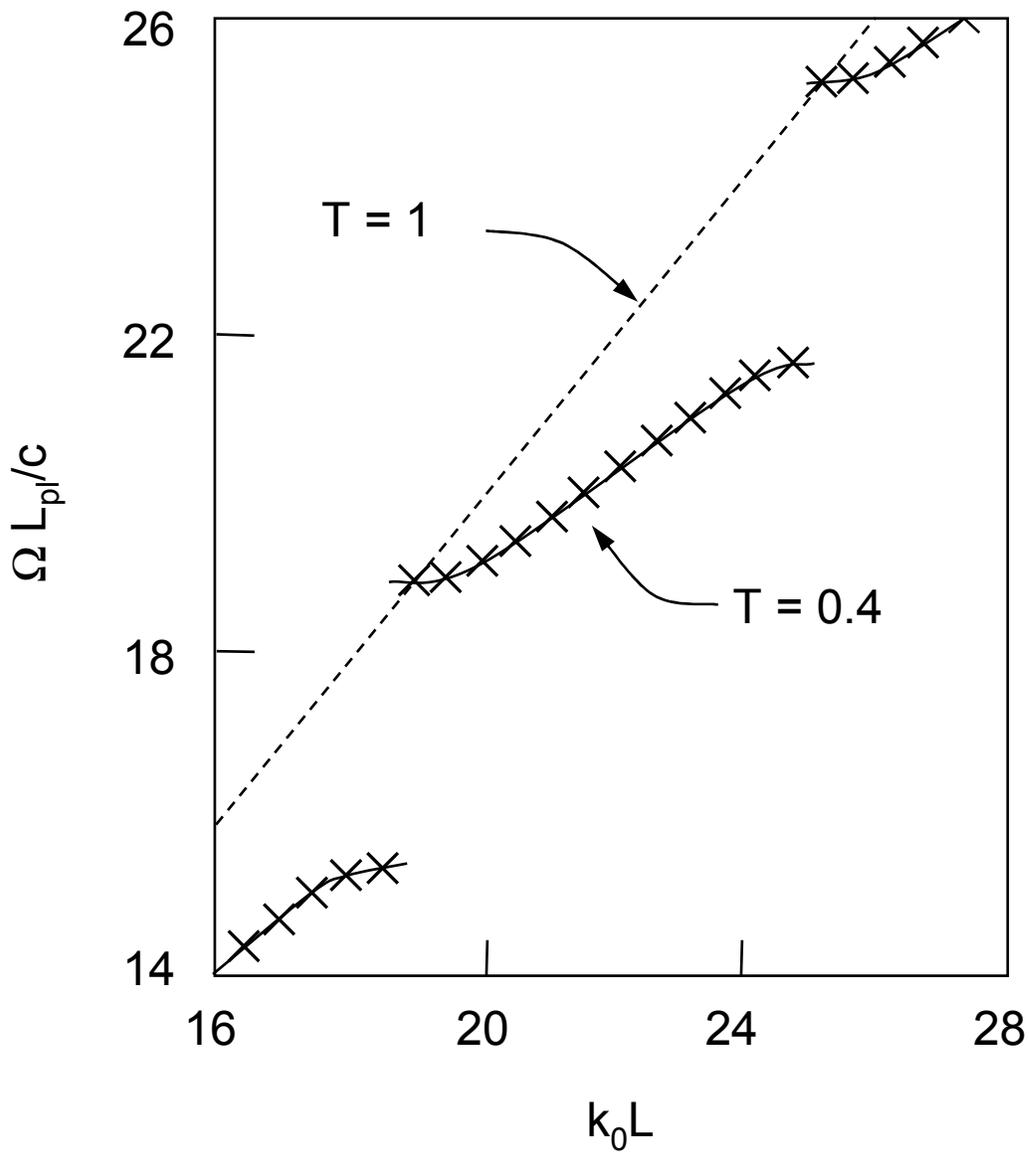
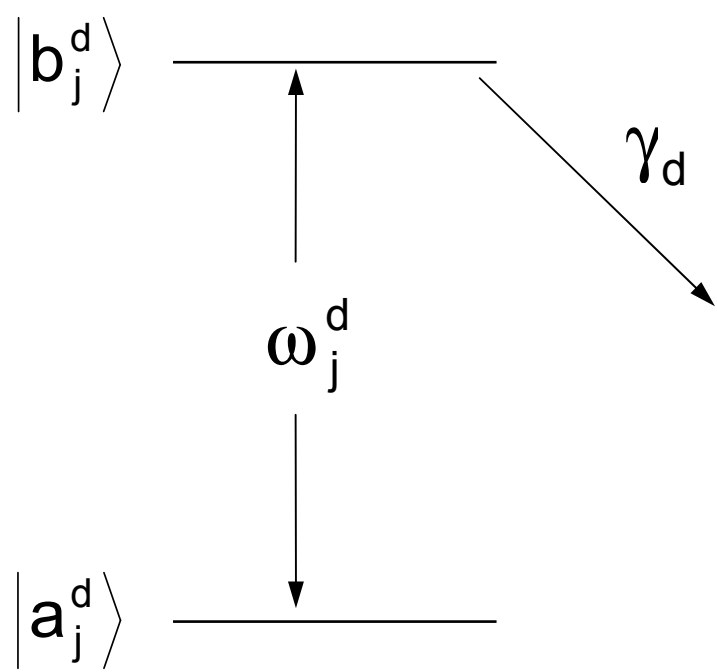
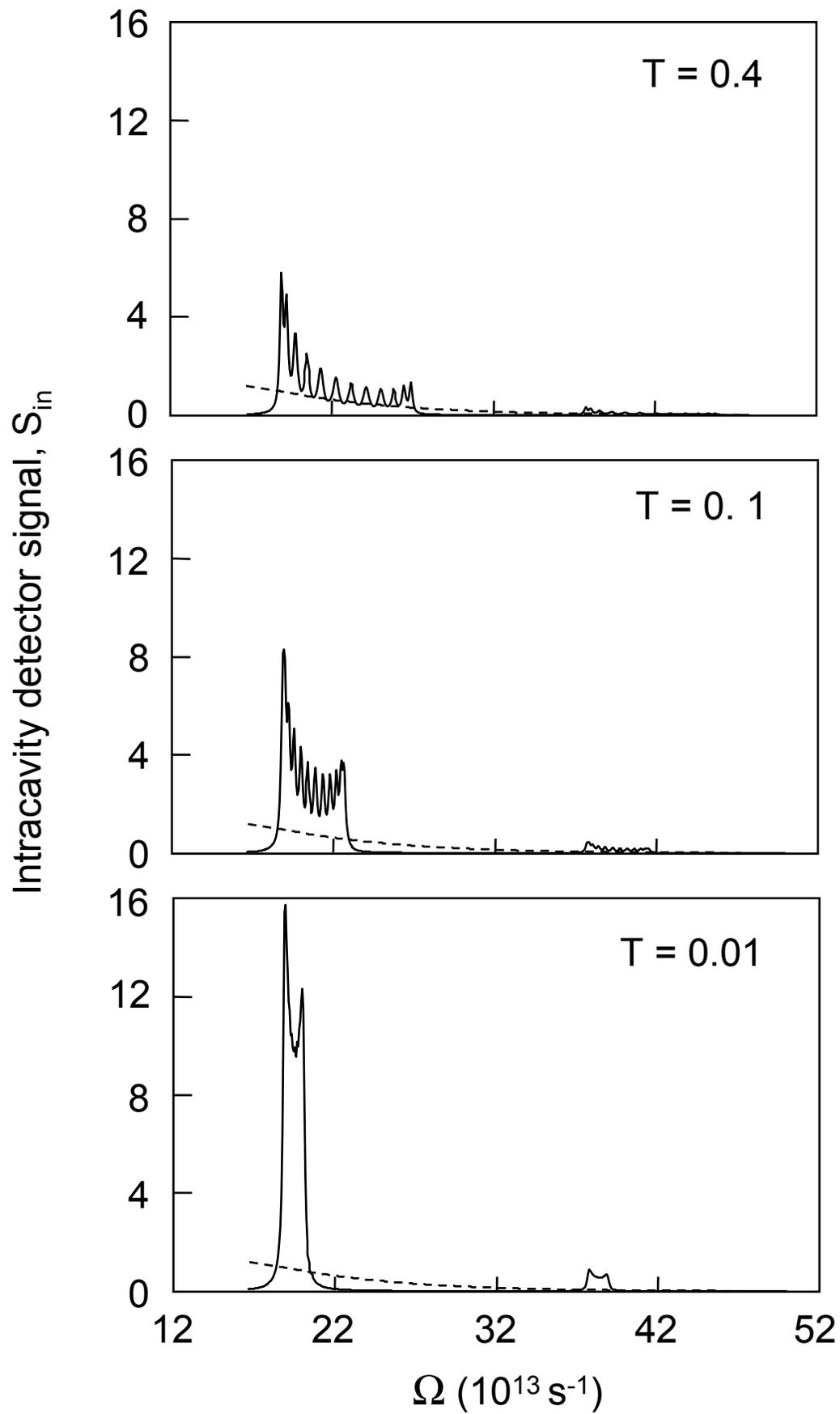
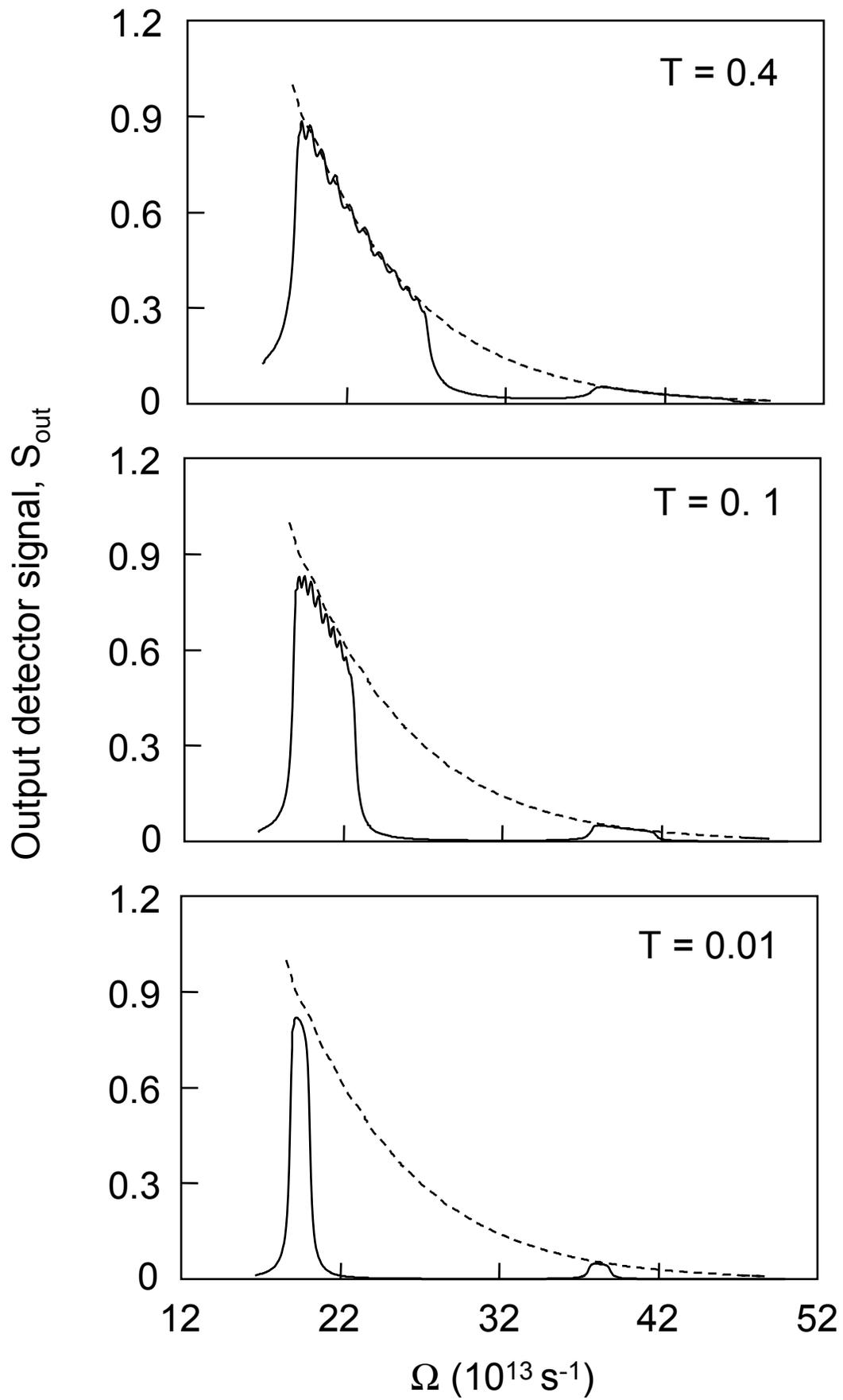
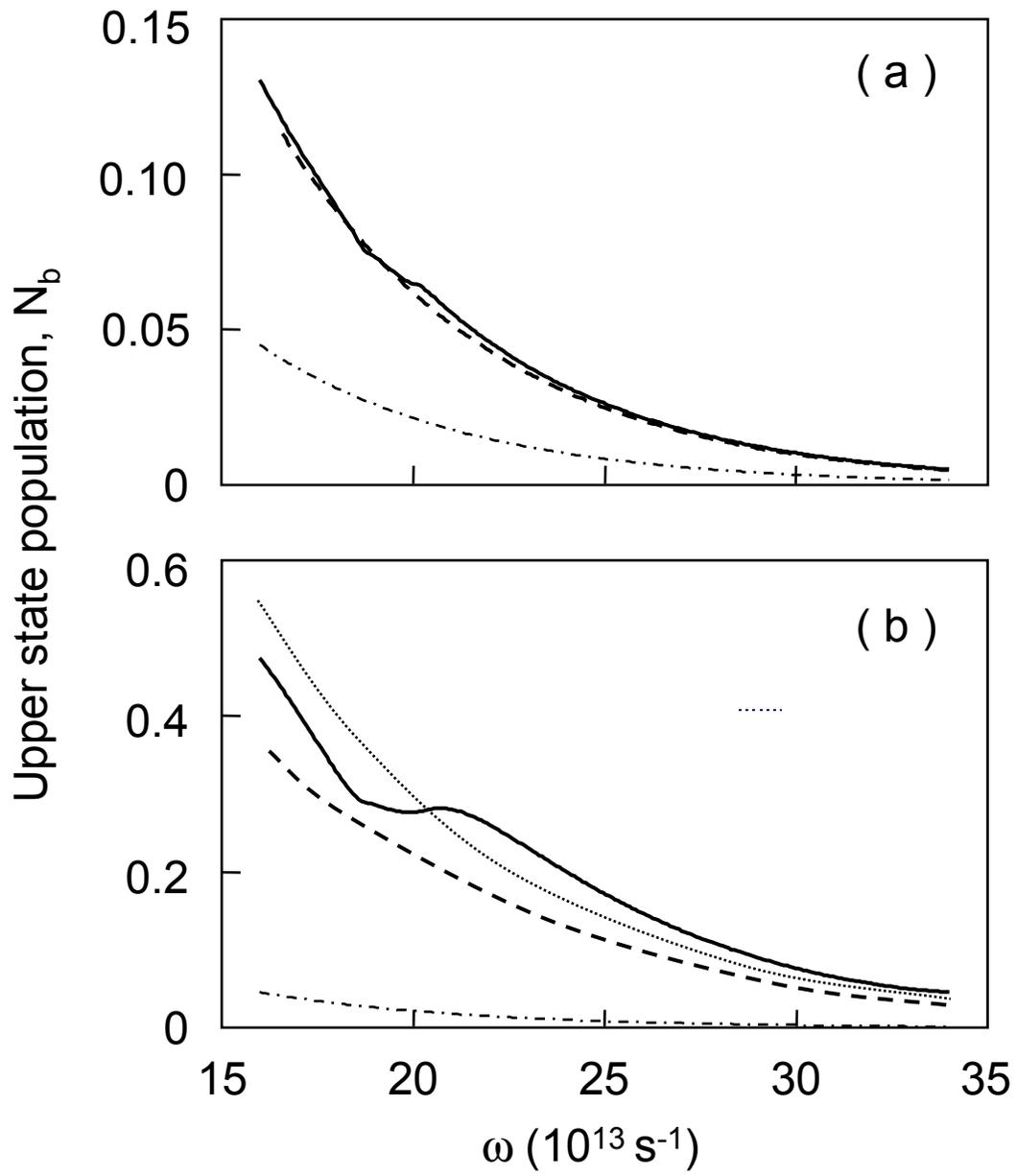
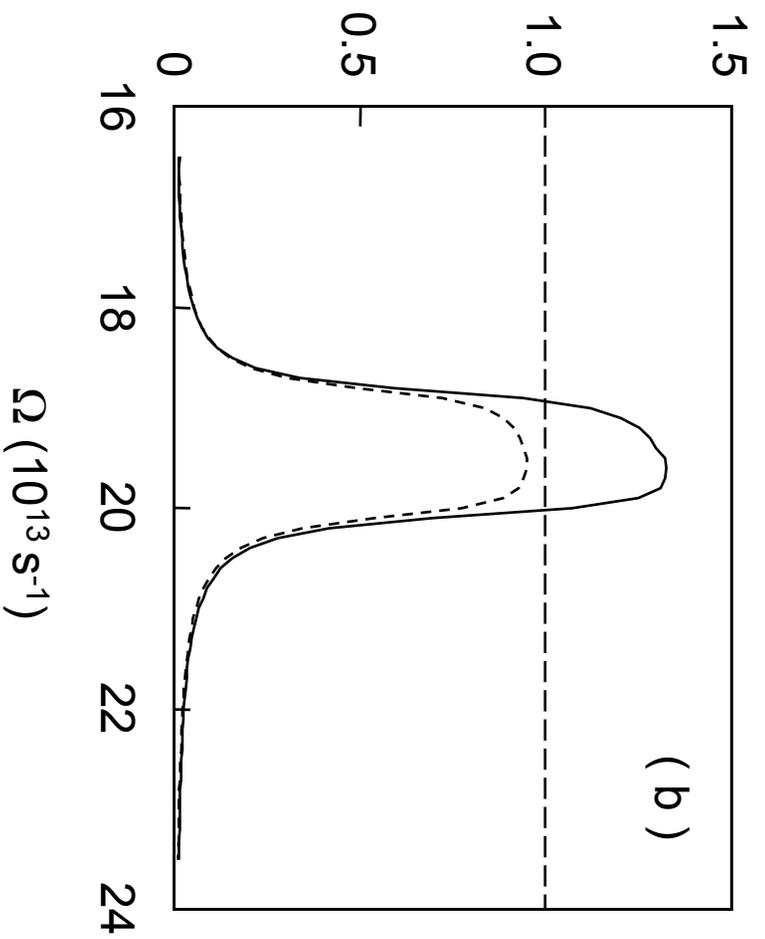
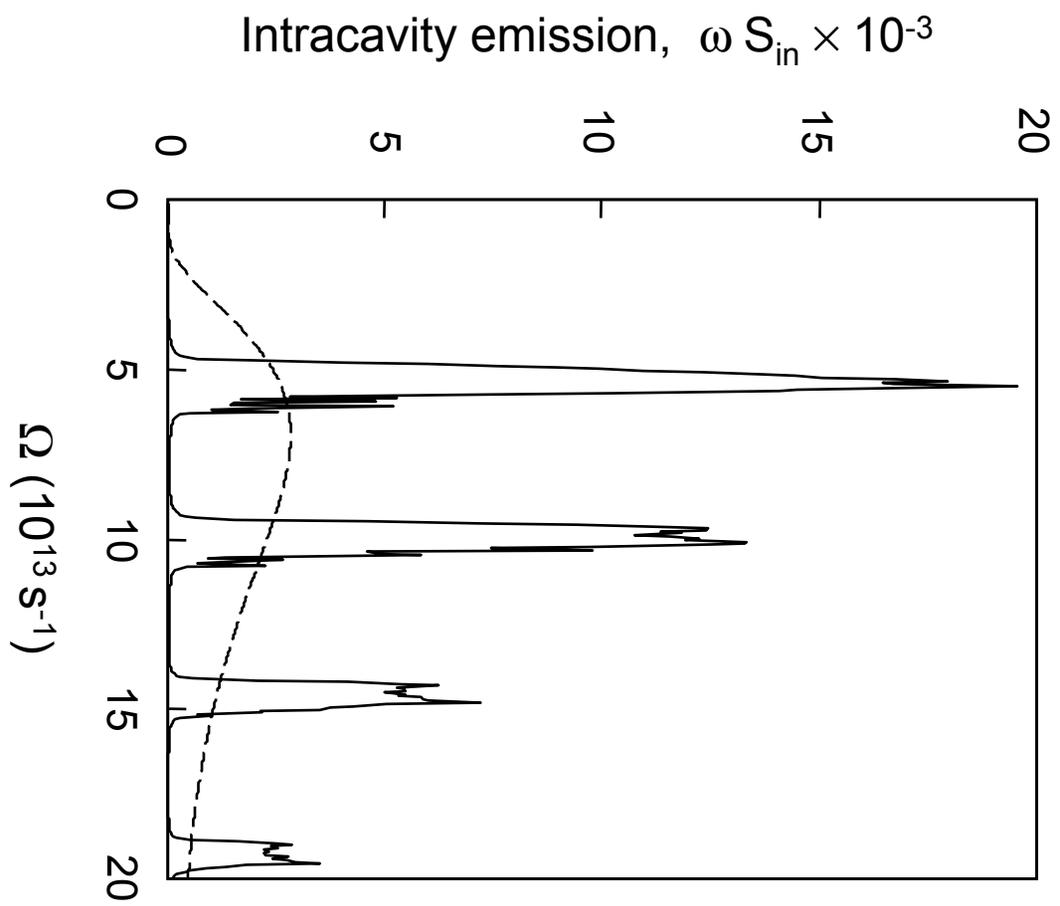
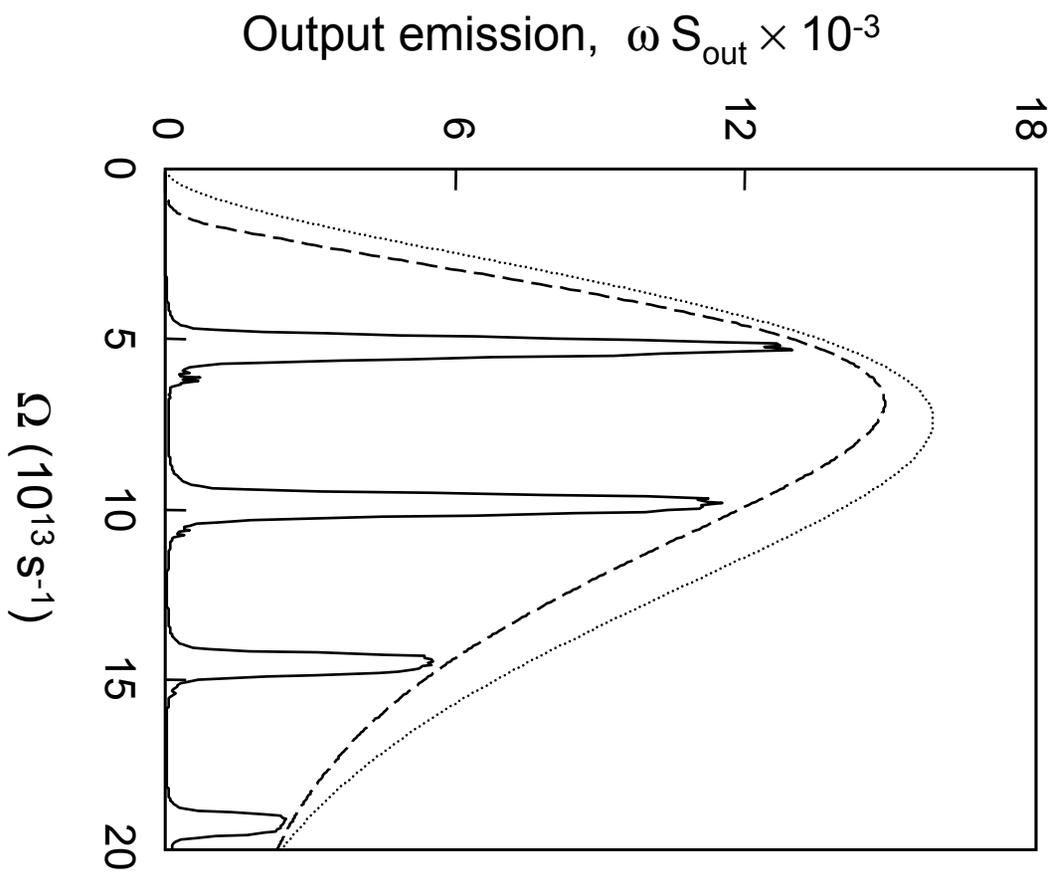
\end{document}